\documentstyle[11pt,aspconf,epsf]{article}
\def\edcomment#1{\iffalse\marginpar{\raggedright\sl#1\/}\else\relax\fi}
\reversemarginpar

\begin{document}

\title{A Unified View of How the Study of Emission Lines Furthers our Knowledge
of AGN}

\author{Julian H. Krolik}
\affil{Department of Physics and Astronomy, Johns Hopkins University,
Baltimore MD 21218, USA}

\begin{abstract}
   A global view is given of how emission lines have been, and may in the
future, be used to enhance our understanding of AGN.  Lines from the microwave
to the X-ray bands all contribute.  Although we have a deep understanding
of the physical processes by which line photons are generated, when the
circumstances of line emission are complicated, models become too unreliable
to provide strong inferences about the rest of the AGN system.  At present,
the lines which appear the most promising for helping answer the most important
questions about AGN are the 22~GHz H$_2$O rotational transition and the 6.4~keV
Fe K$\alpha$ fluorescence.
\end{abstract}

\section{Overview}

    We have seen many presentations in this conference illustrating
the enormous richness of emission line phenomenology in AGN, as well
as the great progress which has been made in understanding their excitation
mechanisms and locating their excitation regions.  In this talk I wish
to step back for a moment and ask what the study of emission lines has
gained us with respect to answering the most fundamental questions about
AGN.  Innumerable proposals requesting support for AGN emission line work
have been justified by the assertion that these lines may, much as emission
lines from fusion plasmas do, serve as diagnostics of the entire system.  I
wish to examine to what extent this assertion has been borne out so far,
and evaluate the prospects for success of this program in the foreseeable
future.

   The natural place to begin is to list the most important questions
of AGN studies.   To my mind, there are six that stand out clearly as being
truly fundamental:

\vskip 0.1in

$\bullet$ 1) What is the power source?  Is it really accretion onto a massive
black hole?

$\bullet$ 2) What triggers nuclear activity in galaxies?  Put another way, what
regulates the accretion flow?  Why was powerful nuclear activity so much
more common at $z \simeq 2$ -- 3 than it is today?

$\bullet$ 3) Why is the continuum emission so broad-band?  What balances the
emission so that (to order of magnitude) the flux is constant per logarithmic
frequency from the mid-infrared through hard X-rays?

$\bullet$ 4) How are jets accelerated and collimated?  Why do they exist
at all?

$\bullet$ 5) How can we relate the different varieties of AGN?
What controls the
choices in type of activity?  For example, what creates the correlation between
host morphology and radio (i.e. jet) power?

$\bullet$ 6) Does the existence of an active nucleus affect the evolution of
the host?

\vskip 0.1in

\noindent Despite more than thirty years of intense activity in this field,
most of these questions remain almost as open today as they were decades
ago.   We do have partial answers to questions 1
and 5, and emission line studies have contributed significantly to providing
those answers, but we have made little headway with the others.
The central issue before us is how emission line studies
may be used to provide more complete answers to all these questions.

   My review of the contributions of emission line work will be ordered
from the outside in, moving from the narrow line region inward toward the
black hole's event horizon.

\section{Narrow Lines}

   Historically most of the effort devoted to analyzing narrow emission lines
has been devoted to inferring the physical conditions in which they are
excited.  However, those details have told us little about other, more
important issues.  It is, rather, the shape of the region from which they
are emitted that has revealed the greatest information.  As reviewed, for
example, by Wilson (1994), in a great many Seyfert galaxies the narrow
line region is a double cone whose apex coincides with the galactic nucleus.
The edges of these cones are often so sharply defined and so straight that
we can imagine hardly any other explanation for them but the projection of
light rays (a particularly clear example is displayed in Figure 1).
That is, the existence of these cones tells us that there is
a sharp-edged toroidal collimating structure close to the nucleus which permits
ionizing photons to escape only within a certain angle of a preferred axis.

\begin{figure}
\plotfiddle{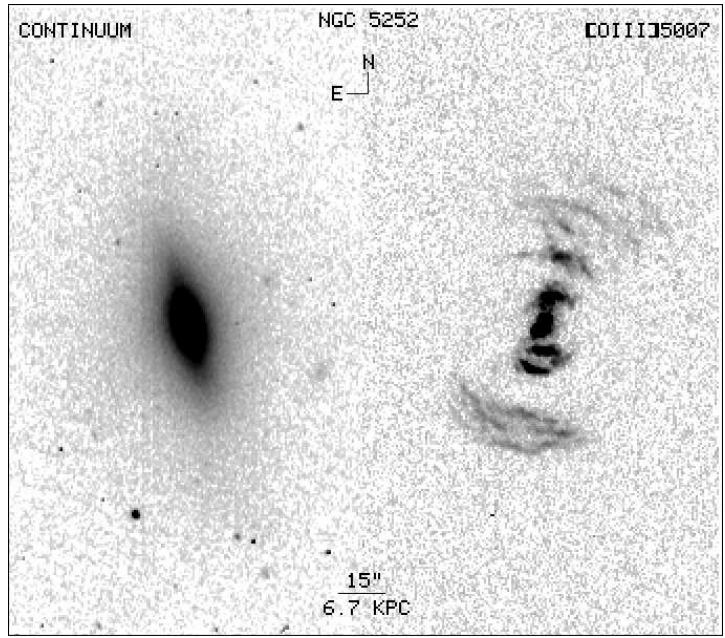}{4.0in}{0}{50}{50}{-155}{-80}
\caption{The ``ionization cone" in the type 2 Seyfert galaxy NGC 5252 (courtesy
Z. Tsvetanov).  The left-hand panel shows a continuum image of the galaxy,
the right-hand an image in a filter centered on red-shifted [OIII]~5007.
The line emission is entirely within a sharp-edged double cone canted at
an oblique angle with respect to the galactic axis.}
\end{figure}

   Several important conclusions follow immediately.  The first is that
AGN appear very different, depending on one's viewing angle.  If an
observer's line of sight lies within the favored cone, a strong ionizing
continuum can be seen; if outside, the ionizing continuum (and likely the
non-ionizing optical continuum also) will appear much weaker.  Thus, there
must be categories of AGN which are separated by empirical classification
schemes but are intrinsically the same.  Some of the now almost
universal support for the unification of type 1 and type 2 Seyfert galaxies (Antonucci 1993) stems from these observations of conical narrow emission line
regions.

    Another conclusion is that the surprisingly weak impact AGN appear to
have on their hosts is due in part to the collimation of ionizing photons
so close to the nucleus.  If the absorbing structure obscures $\simeq 3/4$
of solid angle (as indicated by both the opening angles of the cones
and the statistics of type 1 and type 2 Seyferts), then most of the host galaxy
is protected from exposure to the AGN's ionizing continuum.

    We are beginning to have some grounds for hope that narrow emission
lines may also help us learn about the dynamics of
the accretion flow at distances $\sim 100$ -- 1000~pc from the nucleus.
Nelson \& Whittle (1996) have discovered a strong correlation between the
integrated width of the narrow lines and the velocity dispersion in their
hosts' bulges.  This correlation strongly suggests that narrow line dynamics
are dominated by gravity, although the spatial separation between the narrow
line region and the portion of the bulge where the stellar dispersions are
generally measured makes the nature of this connection somewhat obscure.
Detailed images of the line profiles, as provided, for example, by
Fabry-Perot spectroscopy (Cecil et al. 1990), have the potential
to reveal more about the nature of the forces directing gas towards the center.

    Care will be necessary to successfully carry out this program, however.
If one wishes to interpret gas kinematics as solely due to gravity, it is
vital to omit those regions for which the character of the emission lines
suggests shock excitation rather than photoionization.  Contrasting the
gas's kinematics with the kinematics of adjacent stars should go a long
way towards helping us make these distinctions.

\section{Reflected Lines}

    In a classic paper, Antonucci \& Miller (1985) demonstrated that
broad Balmer emission lines, while invisible in the total flux of the classic
type 2 Seyfert galaxy NGC~1068, are as strong in its polarized spectrum as
in any type 1 Seyfert.  Based on this finding, they inferred that the nucleus
of NGC~1068 (and by the astronomical version of mathematical induction,
the nuclei of most,
if not all, other type 2 Seyfert galaxies) are in fact identical to the
nuclei of type 1 Seyferts but for the accident of very optically thick
obscuration that blocks our line of sight.  The simplest geometry for
this obscuration is toroidal, as confirmed by the conical shape of
the narrow line region. We see the
intrinsic spectrum of the nucleus only because a small fraction of the light
is reflected our way by electrons within the illuminated cone but above
the obscuration, thus having unencumbered lines of sight to
both the nucleus and our telescopes.  The polarization created by the
asymmetric disposition of these electrons on the sky allows us to identify
the light as coming from the nucleus.  Clearly, the presence of strong
broad emission lines in the polarization spectrum served in this case as
a marker of nuclear reflected light.

    Many others (again see the review by Antonucci 1993) have told the
story of how polarized emission lines showed us the path to unification of
the different Seyfert types, as well as (more speculatively) FR2
radio galaxies and radio-loud quasars.  Here I wish to emphasize a less
well-known way in which emission lines from the mirror tell us more about
AGN.  Because the primary scattering mechanism is electron scattering,
the profile of any reflected emission line with line-center frequency
$\nu_o$ is both broadened [by $\delta \nu \sim \nu_o (kT_e/m_e c^2)^{1/2}$]
and shifted [by $\delta \nu = \nu_o (\vec v/c) \cdot (\hat n_f - \hat n_i)$],
where $T_e$ is the electron temperature, $\vec v$ the electrons' bulk
velocity, and $\hat n_{i,f}$ are the initial and final directions of the
scattered photons.  Thus, the detailed profile properties of the polarized
emission lines function as diagnostics of the physical properties of
the mirror gas.  For example, in NGC~1068, Miller, Goodrich, \& Mathews
(1991) were able to show that $T_e \simeq 3 \times 10^5$~K, and that the
H$\beta$ line is shifted 600~km~s$^{-1}$ from systemic in the sense that
the scattering medium is moving outward.

   X-ray spectroscopy of the mirror gas has the potential to tell us even
more.  When our line of sight is in the equatorial plane of the obscuration,
we see the mirror gas by virtue of its intrinsic emission and its reflection
of the nuclear light (the specific spectrum predicted by a particular model
is shown in Figure 2).   When this gas is warm (i.e. $T \sim 10^5$ -- $10^6$~K
as in NGC~1068), electron scattering is the only scattering mechanism
effective in the optical and ultraviolet bands, but numerous resonance lines
enhance the gas's albedo in the soft X-ray band (Band et al. 1990;
Krolik \& Kriss 1995).  Exactly which lines are present, and at what
strength, depends on details of the physical conditions.  In particular,
we have good reasons, both observational (the line shift seen in NGC 1068
described in the previous section) and theoretical (Krolik \& Begelman 1986,
Balsara \& Krolik 1993) to believe that this gas is flowing outward
supersonically.  In such a state, the gas absorbs somewhat more heat from
radiation than it emits, so its temperature is not exactly the one
given by thermal balance at a fixed ionization parameter.  Consequently,
to specify its ionization state one must define both the ionization
parameter and the temperature.   Dynamical models (e.g. the two papers
just cited) suggest that $\Xi$ (the ionization parameter defined as
a ratio between radiation and gas pressure) is $\sim 10$ and the temperature
is between $10^5$ and $10^6$~K.  That is, the gas is slightly to the net
heating side of the equilibrium curve near the region of that curve
where it is marginally stable with respect to isobaric perturbations.

   In Figure 2 the spectrum has been smoothed with a Gaussian filter having
a fractional frequency width of 0.05 in order to illustrate the richness
of the structure which might be seen when X-ray spectrographs are able to
obtain both resolution on this scale, and adequate signal/noise to make use
of it.  The features seen are actually blends of numerous lines, both
intrinsic and reflected, whose shapes and strengths depend strongly on
the state of motion of the mirror gas.  Supersonic motion has an especially
powerful effect on the reflected lines because it can move them from
the flat part of the curve-of-growth to the linear part: the ratio between
the velocity width of a line formed in a supersonic flow and the velocity
width that line would have in a static gas is $\Delta v_{wind}/
\Delta v_{thermal} \sim A^{1/2} {\cal M}$, where $A$ is the atomic mass of the
ion and ${\cal M}$ is the Mach number of the flow.  For those ions
with resonance lines in the soft X-ray band, $A$ is typically a few tens,
so a supersonic velocity gradient can widen their lines by an order of
magnitude.

\begin{figure}
\plotfiddle{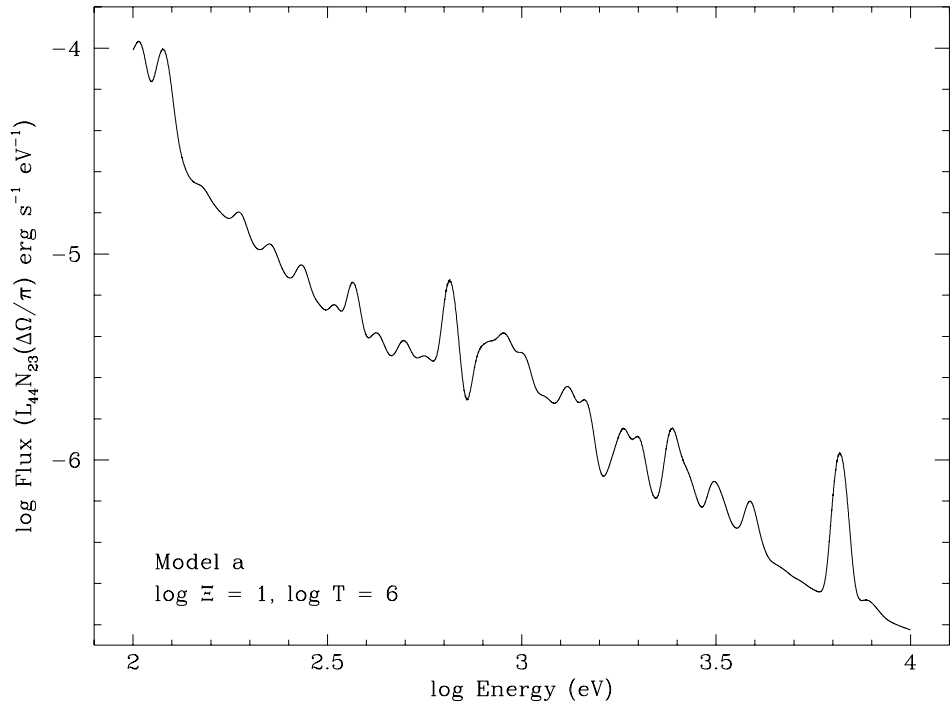}{4.0in}{270}{50}{50}{-200}{+300}
\caption{The EUV through X-ray spectrum of an obscured AGN, including both
intrinsic emission from the mirror and resonance scattering, as predicted by a
plausible model for its physical conditions (Krolik \& Kriss 1995).
The velocity distribution was assumed to be a Gaussian with a characteristic
width equal to the sound speed.}
\end{figure}

    A polar view of the mirror (and nucleus) provides us with a different
view that is potentially equally informative in a complementary way.  In
this case the features are more likely to be absorption than emission.
Because the gas is only partially ionized, photons are removed from the
nuclear continuum both by photoionization (true absorption) and by
resonance scattering (as just discussed).  Because the mirror covers at most
$\simeq 1/4$ of the solid angle
around the nucleus, reemission by radiative recombination and
scattering from other lines of sight into ours replace only
a small part of the photons removed by absorption and scattering.
Viewed down the axis, then, the mirror acts as an absorption filter in
the soft X-ray band.

     It is fair to ask whether these predicted absorption features have ever
been seen.  If the mirror covers most of the solid angle left open by the
obscuration, they must occur in many, if not most, type 1 Seyfert galaxies.
Estimates of the fraction of the nuclear light reflected (e.g. Miller et al.
1991) indicate a column density $\sim 10^{22}$ -- $10^{23}$~cm$^{-2}$, and,
as we have seen, other measurements and theoretical arguments
suggest a temperature in the range
$10^5$ -- $10^6$~K and an ionization parameter slightly larger than the
value at which marginally stable thermal equilibria are found.
When radiative equilibrium models are fit to warm
absorber spectra, very similar column densities and temperatures are
found, and the ionization parameter is estimated to be very close to
the marginally stable value (e.g. Netzer 1997).
The mirror is, therefore, an excellent candidate to {\it be} the warm absorber.  X-ray absorption line studies may then provide another vehicle
for learning about the character of the reflection region.

\section{The Obscuring Torus}

   Emission lines from the obscuring material are also useful.  Covering
$3\pi$ of solid angle, it should be a source of substantial Fe K$\alpha$
emission if its column density is $\sim 10^{24}$~cm$^{-2}$ or more (Krolik
et al. 1994; Ghisellini et al. 1994).  Since much of the obscuring
matter is located
1 -- 10~pc from the nucleus (for Seyfert-scale luminosities), the flux in
Fe K$\alpha$ produced in the obscuring torus can only change on timescales
of at least a few years.  Therefore, very long duration (10~yr or
more) reverberation mapping experiments might be able to map out the
location of K$\alpha$-emitting material in much the same way as shorter duration
reverberation mapping has given us much information about the location of
broad emission line gas (Maoz 1997).  In at least one object (NGC 2992),
incomplete time sampling shows line and continuum variations of just
the sort that could be used for this kind of experiment (Weaver et al. 1996).

   The obscuring torus houses a still more powerful emission line diagnostic
in the form of 22~GHz H$_2$O masers (Braatz et al. 1996).  Because
interferometric techniques are so well developed at radio frequencies, and
the angular scale of these masers is comparatively great (relative to
other AGN structures), we are able to obtain genuine {\it images} of their
location, without any sort of inferential guesswork.  In the case of
NGC 1068, we can see them outlining the inner edge of the obscuring torus
(Greenhill 1997)!

   Combining direct positional information with line of sight velocities,
we can conduct detailed investigations of the torus dynamics using these
maser spots.  When, as in the
case of NGC 4258 (Miyoshi et al. 1995), those dynamics are simple,
it is possible to make very robust and precise inferences about the overall
gravitational potential.  Because the maser spots in that object
are very closely confined to a thin (albeit slightly warped) plane, the
spots travel
in orbits that are very close to circular, and the magnitude of their
orbital speed scales $\propto r^{-1/2}$, it is very hard to escape the
conclusion that their motions are simple tracers of a point-mass potential
created by a central mass of $3.6 \times 10^7 M_{\odot}$ whose size
must be rather smaller than 0.1~pc.

   On the other hand, there are also examples, such as NGC~1068 (Greenhill
1997), in which the dynamics are complicated.  Interesting inferences may
still be made, although the arguments are much less clean.
In that case, there is a
very large velocity dispersion at the innermost radii, and the upper envelope
of the line of sight velocity scales much more slowly with radius than
$r^{-1/2}$; in fact, it is more like $r^{-1/20}$.  If one nonetheless
makes the simplistic assumption that the motions are due primarily
to gravitational forces, the orbital speed at the innermost radius
may be used to yield an order of magnitude
indicator of the central mass, i.e. $M \sim rv^2/G \sim
10^7 M_{\odot}$.  Compared to the bolometric luminosity of
$10^{45}$~erg~s$^{-1}$ (Pier et al. 1994), this means that the
luminosity relative to Eddington is $\sim 0.6$.  Inside the torus, the
spectrum peaks in the mid-infrared where, for a normal dust-to-gas ratio,
the opacity per unit mass is about ten times Thomson.  Since the radiation
force is the flux times the opacity, these
observations demonstrate that radiation force {\it must} be important to
the dynamics of the torus, as predicted by Pier \& Krolik (1992).

   These arguments illustrate an important general principle: Simple
sub-systems (e.g. the maser spots in NGC~4258) can be so thoroughly understood
that they may be reliably used for inferences about the larger system in
which they are embedded.  But complicated systems (e.g. the maser
spots in NGC~1068) are never well enough understood that they may be used
as springboards for inferences about larger issues.  Inevitably we find
ourselves fully occupied refining what we know about their {\it intrinsic}
properties.  That activity may be fascinating and well worth the effort, but
it does not lead to information bearing on larger questions.

\section{Broad Lines}

    Having said that, we are now equipped to consider the role of broad
emission line studies in the context of the broad sweep of AGN work.  These
lines offer a tremendous wealth and quality of data.  In any one object
there are numerous different lines; the flux and profile of each one may
be measured; and they vary in time in a way which is correlated with the
continuum.  At the same time, because the atomic physics of optical and
ultraviolet line emission is well understood, it is possible to make
very detailed calculations modelling their excitation and have some fair
confidence that the results are meaningful.  By comparing
the relative strengths of different lines one may infer the ionizing flux
and pressure of the region where the lines are made; using the methods
of reverberation mapping one may also constrain the distance from the
central continuum source to the line-emitting region, thereby fixing
all the unknowns of the problem.  The pressure in the line-emitting
regions might be taken as a measure of the pressure in the surrounding
gas, which, if we were lucky, might be identified with the accretion flow
through that range of radii.  The line profiles give measures of
bulk velocities; again, with good fortune, we might hope that these velocities
indicate at least something about the velocities of surrounding matter.
On this basis, many people
have long expected that the study of AGN broad emission lines would
cast a bright light on the field as a whole.

   Unfortunately, there are numerous obstacles that may prevent us from
ever reaching these goals.  Reverberation mapping experiments (again see
the review by Maoz 1997) have shown that the broad line region is a very
complex place.  The material responsible for individual lines
is spread over at least an order of magnitude
dynamic range in radius, and maybe more (Done \& Krolik 1996).  The
ranges of radius corresponding to different lines are systematically offset
from each other by sizable amounts (e.g. Clavel et al. 1991,
Krolik et al. 1991).  And the
whole structure may change in timescales as short as a few years (Peterson
\& Wanders 1996).

  Moreover, even obtaining a believable reverberation map is subtler than
might be apparent at first sight.  The effective signal to noise ratio
for a monitoring experiment is {\it not} the ratio of mean flux to rms
error, but rather the ratio of rms flux change (that is, compared
to the long-time mean of the flux)
to rms error.  In most monitoring experiments so far this ratio has been
only a few; even in the best (measurements of the total flux in the CIV~1549
line in the {\it HST} campaign on NGC~5548 reported by Korista et al. 1995)
it rose only to $\simeq 40$.  Blending is frequently a problem, corrupting
otherwise easily-measured lines.  It is particularly frustrating that
NV~1240 almost always strongly contaminates Ly$\alpha$.

   Even when the data are very clean, inversion of the line and continuum
time series to obtain the response function {\it requires} insertion of
some {\it a priori} information in order to achieve numerical stability.
This automatically means that {\it all} response functions are model-dependent
on some level (Figure 3).  It is possible to make this model-dependence
relatively benign, but it is always present.  As Figure 3 makes plain,
while there are certain features that are robust with respect to changes
in the model assumptions, there are definitely others which are clearly
model-dependent.

\begin{figure}
\plotfiddle{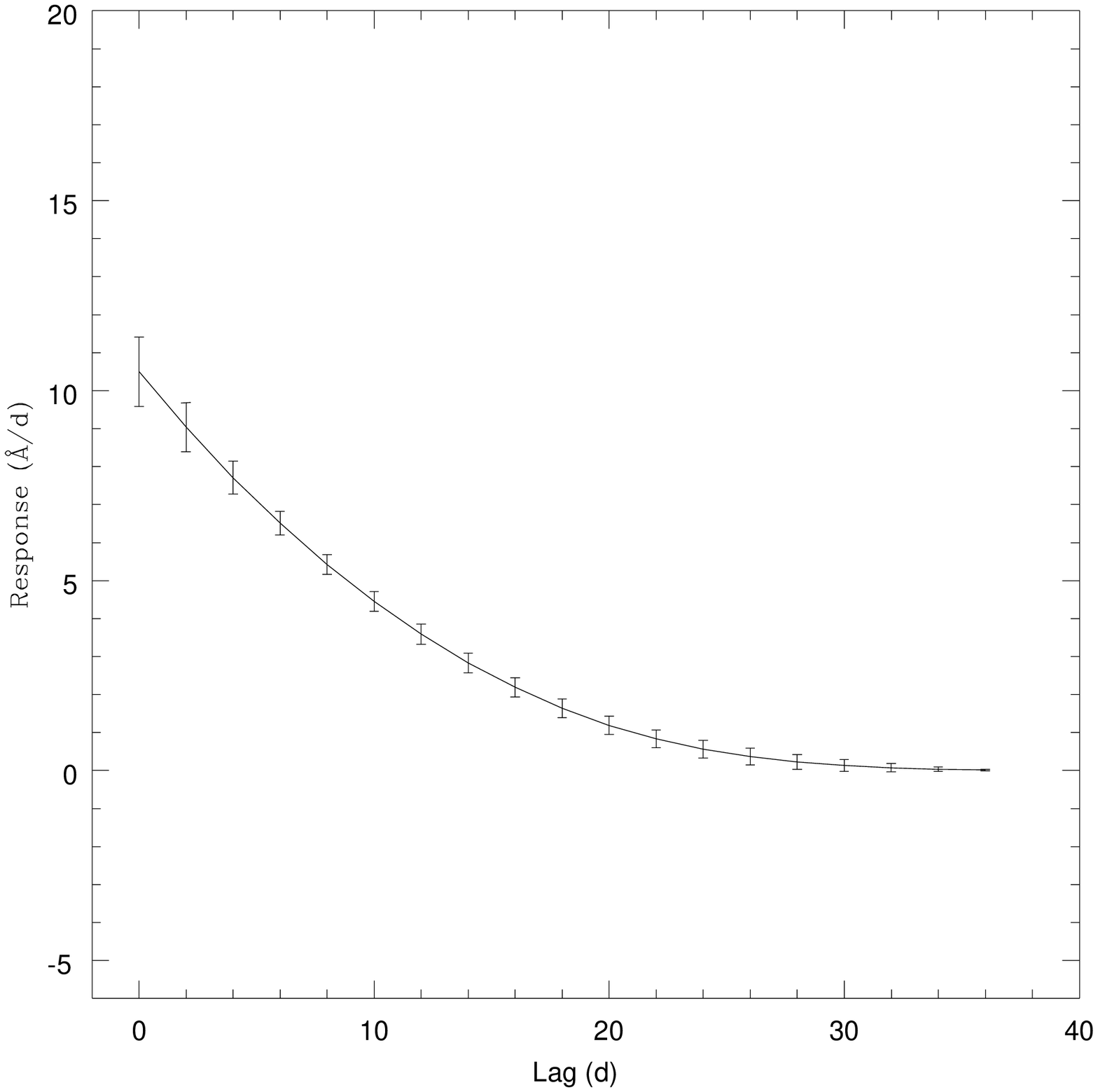}{0.0in}{0}{35}{35}{-220}{-243}
\plotfiddle{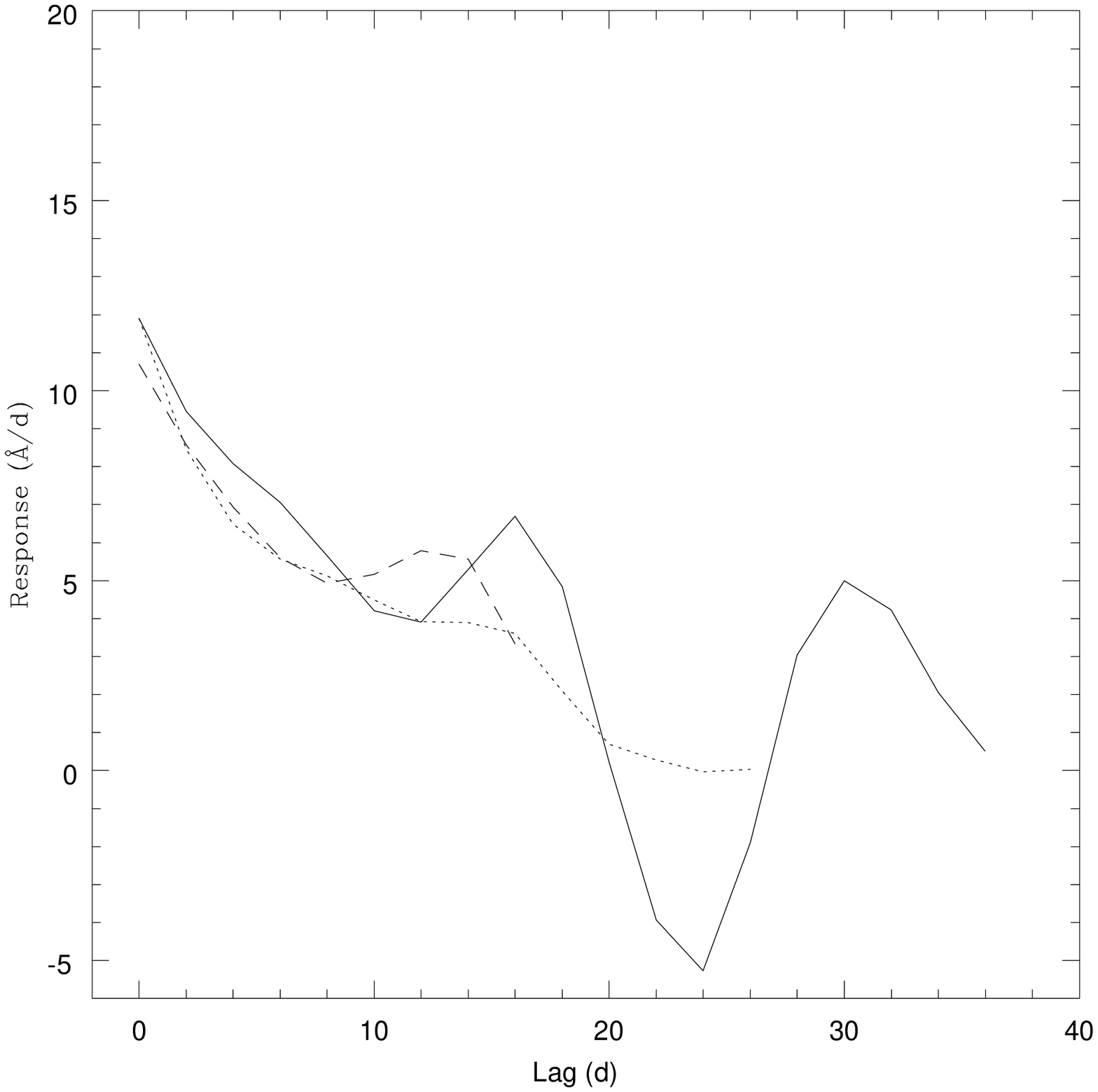}{2.5in}{0}{35}{35}{-10}{-50}
\caption{The model-dependence of response functions.  The left-hand
panel shows a solution for the total CIV flux data from the NGC 5548
{\it HST} campaign which was tuned to yield the smoothest response function
consistent with the data.  The right-hand panel shows response functions
obtained from the {\it same} data, but for a range of maximum lengths,
and reducing the relative weight given the smoothness constraint so that
it has no more weight than the data.}
\end{figure}

   To go from the 1-d map given by the response function to a true 3-d
picture of the emission line region one inserts far greater model-dependence.
A response function is no more than a projection of the marginal response
of line emissivity onto the surfaces of constant delay {\it relative to
our line of sight}.  We certainly cannot expect our line of sight to be
a symmetry axis for any but a tiny minority of AGN, so we must make some
choice about the real symmetry in order to transform the response function
into a 3-d description.  This procedure is obviously highly non-unique.

   The potential pay-off from reverberation mapping becomes still greater
when separate maps are made not just for different emission lines, but
for different portions of the lines, separated by line-of-sight velocity.
There is no change in the technical issues; the only respect in which
velocity-resolved reverberation mapping is more difficult than integrated
flux reverberation mapping is the reduction in signal/noise due to
chopping up each line into several segments.  What makes the scientific
yield so much more interesting is that the solution provides a constraint
on the {\it kinematics} of the line-emitting gas, which we may hope will
ultimately provide clues about their {\it dynamics}.

   Unfortunately, the remarks made above about non-uniqueness
are underlined by the current state-of-the-art in velocity-resolved
reverberation mapping.  Four dramatically different models for the NGC~5548
CIV~1549 response have been put forward, all claiming qualitative
consistentency with the data (in order of publication, Wanders et al. 1995;
Done \& Krolik 1996; Chiang \& Murray 1996; Bottorff et al. 1997).
The first argues that the velocities are all random, but that a ``shelf"
exists in the total flux response similar to that appearing in the solid
curve of Figure 3b.  They interpret this shelf as due to strong anisotropy
of the ionizing continuum, rejecting the numerous other possible interpretations
of such a feature.  The second finds no evidence against spherical
symmetry and isotropy in the emission.  On the other hand, while their
analysis shows (like that of the first paper) that the line's red and
blue cores respond similarly, they differ from Wanders et al. in
discovering that the red wing responds
significantly more quickly than the other line segments, especially the
blue wing.  On this basis they suggest that the line motions are predominantly
random, but there is a net radial infall.  The third posits a rotating wind
driven off a flat accretion disk by radiation pressure, and argues that
the anisotropy of line emission in this sort of situation creates leading
response in the red wing despite the outflow.  The fourth also favors
a disk wind, but driven by rotating magnetic fields.

   The only hope we have to break this degeneracy is to impose the twin
thresholds of physical self-consistency and quantitative (e.g. as measured
via a statistic such as $\chi^2$) consistency with the data.
None of the above models
passes both tests fully: the first two are physically {\it ad hoc}, while
the second two are based on incompletely worked out dynamical models; only
the second has even computed the $\chi^2$ for its predictions. 

   Let us suppose that over the next few years we are able to arrive at a
generally-accepted picture of the kinematics of the broad line region.  How
much will we able to learn then?  For a number of years, it has been commonplace
in the literature to assume that the dynamics of the line-emitting material
are dominated by the gravitational force of a massive central object, and
that its mass could be estimated simply by computing $r v^2/G$.  Unfortunately,
these statements have as yet little or no justification.

    First of all, we do not yet have any clear indication regarding which forces
dominate the broad emission line gas.  In any gas optically thick over
much of the continuum bandwidth, the ratio between the mean outward radiation
force and the inward gravitational force due to the central object is
$${F_r \over F_g} \simeq {L/L_E \over \tau_T},$$
where $L/L_E$ is the luminosity in Eddington units and $\tau_T$ is the optical
thickness of the gas in Thomson units.  We do not know with any confidence
either of the two dimensionless numbers taken in ratio on the right hand
side of this equation.  Both have been suggested to lie anywhere in the range
$10^{-3}$ to $\sim 1$.  We can hardly state with confidence, therefore, that
radiation force is clearly negligible relative to gravity.

   Other forces may also play a role.  While we have good reason to believe
that the broad emission line gas occupies only a very small fraction of the
volume of its region, we have no idea what sort of gas fills the rest of
the volume.  Friction against this other gas could be very important to the
emission line gas.  Similarly, we have even less idea about the state of
magnetic fields running through the region, so we have no sensible way to
even begin to limit $(\nabla \times \vec B) \times \vec B$ relative to
gravity (see, for example, Rees 1987 or Emmering et al. 1992).

Indeed, contrary to what is often said, $r v^2/G$ does not even provide a
limit.  If
other forces dominate gravity, the number it gives for the central mass
is an overestimate; if other forces are comparable to gravity and partially
cancel it, this number gives an underestimate. 

   Even if in the future we are able to eliminate significant roles for forces
other than gravity, there are several complications that stand between us and
using broad line kinematics to measure the mass of the central black hole.
First of all, it is essential to have a complete picture of the kinematics,
not just measures of the characteristic size and velocity.
As we have just finished discussing,
one of the most important results of the reverberation mapping program has
been to demonstrate that emission line material exists over a very broad
range of radii.  Simply inserting some sort of weighted-mean radius and
(differently) weighted-mean velocity in the formula $rv^2/G$ can
be extremely misleading.  In fact, one way we might some day demonstrate that
the motions are due to the gravitational force of a point-mass would
be to use velocity-resolved reverberation maps to verify that
$v \propto r^{-1/2}$ (intriguingly, Done \& Krolik 1996 suggest that
the scaling of $v$ with $r$ is a good deal slower than this).
Furthermore, although
$r v^2/G$ does provide an order of magnitude estimator of the central mass in
the event that the motions are purely orbital, it is no better than that
without additional information about the shapes of the orbits.
When the orbits are exactly circular, the central mass is exactly $rv^2/G$;
but if the velocities at any given location are distributed randomly
in direction, the central mass is three times greater.

   Finally, if, as has long been hoped, we are to eventually use the broad
emission lines as accretion diagnostics, we must find some way to
relate the gas which emits them to the accretion flow.  If the broad emission
line material is {\it not} moving inward, it cannot itself be taking part
in the accretion process, although it might perhaps be nearby.  Learning how
to draw this relation is unlikely to precede a clear understanding of
the line-emitting matter's dynamics.  On the other hand, if the broad
line material is going in, we might speculate that it comprises the accretion
flow, or is at least a part of it.  To feed a typical Seyfert galaxy with
a luminosity $\sim 10^{44}$~erg~s$^{-1}$, an accretion rate of
$\simeq 0.02 L_{44}(e/0.1)^{-1}$~$M_{\odot}$~yr$^{-1}$ is needed, where
$e$ is the accretion efficiency in rest mass units.  If one interprets
the reverberation mapping of NGC 5548 as showing net inflow, the associated
mass inflow rate is
$\sim 0.001 N_{22} r_{10} v_{1000} [(dC/d\ln r)/0.1] M_{\odot}$~yr$^{-1}$,
where the (not well-determined) column density of the gas has been scaled in
units of $10^{22}$~cm$^{-2}$, the radius in units of 10 lt-d, the velocity
in units of 1000~km~s$^{-1}$ (roughly the best-fit number for the net radial
speed found by Done \& Krolik 1996), and $C$ is the covering factor of the
gas around the central source of radiation.  Thus, even if one accepts
the net inward flow interpretation of the reverberation mapping, the line-emitting stuff may not be more than a small part of the total accretion.
If that is so, we are still faced with the problem of understanding its
dynamics well enough to know how to use it as a tracer of the bulk of
the accretion.

   Broad emission lines, therefore, despite (or perhaps because of) their
phenomenological richness appear to be a prime example of the general
principle enunciated in the previous section.  They are so complicated that
we may never go beyond studying their {\it intrinsic} properties; using them
to answer fundamental questions about the global AGN system may forever be
beyond our reach.

\section{The Accretion Disk}

   Finally we come to the innermost observable portion of any AGN, the
accretion disk.  By far the strongest emission line that we can expect to
be emitted from the accretion disk is the Fe K$\alpha$ fluorescence line,
but numerous other, weaker lines are also possible, mostly with energies
$\sim 1$~keV (e.g. as described in \.Zycki et al. 1994).  Improvements
in both the energy resolution of X-ray spectrometers and their S/N might
eventually allow us to study the weaker lines, but in the meanwhile,
the Fe K$\alpha$ line, which has an equivalent width that is commonly
a few hundred eV, is already very informative (Nandra 1997).  The
discovery, enabled by the ASCA satellite, that this line commonly has
a FWHM between 20,000 and 70,000~km~s$^{-1}$ is a dramatic signal that
relativistic motions are taking place deep inside AGN.  That the profile
is frequently offset to the red also suggests a significant gravitational
redshift, and the plausible supposition that those relativistic motions
are simply orbital motions in a relativistically deep gravitational
potential.

    Moving from qualitative to quantitative results will take a great deal
more effort, however.  Several detailed calculations must be performed in
order to adequately model the K$\alpha$ emission in these circumstances,
and we do not yet know how to carry out every step in these calculations.

First, the line is excited by fluorescence, so this entails a radiation
transfer calculation of modest scope to determine how many K-edge photons
reach how deep into the material.  If the opacity of the material to K-edge
photons were independent of ionization state, this would be a straightforward
calculation.  However, because a significant part of the opacity at energies
above 7.1~keV (the edge energy for neutral Fe) is due to other elements,
one must also calculate the ionization structure self-consistently.  This,
in turn, requires a model of the underlying disk structure.

     Unfortunately,
we do not yet have a satisfactory description for the equilibrium state
of accretion disks around black holes.  The standard Shakura-Sunyaev (1973)
equilibrium is thermally unstable when the accretion rate is more than
$\sim 10^{-3}$ of the Eddington rate (Shakura \& Sunyaev 1976), and we
do not know what other equilibrium replaces it. 
 
    Nonetheless, if one assumes a model for the (un-irradiated)
density and temperature in the disk, it is possible to then compute in
a self-consistent manner its density, temperature, and ionization structure
when irradiated by X-rays (see, for example, Ross \& Fabian 1993 and
\.Zycki et al. 1994).  It is important to know the ionization distribution
of Fe as a function of depth because the energy of the K$\alpha$ photon
depends on ionization state.  The dependence is weak until the atom is
mostly stripped, but can be significant if highly-ionized stages are reached:
the K$\alpha$ line in H-like Fe has $\simeq 8\%$ greater energy than
the line emitted by neutral Fe, a difference comparable to the
frequency shifts expected from relativistic effects.

   After the spectrum of the initial photons is calculated, one must next
find the angular distribution and spectrum with which they emerge from
the disk surface.  Note that the angular distribution with respect to the
system axis may be different from the angular distribution with respect to
the local surface normal if the disk surface is curved.
Some of the line photons may be absorbed by photoionizing lower-Z atoms before
escaping from the disk, others may change direction and energy as a result of
electron scattering.  With the quality of modern X-ray spectroscopy, the
last effect is significant: Compton recoil losses are about 1\% per
scatter; thermal broadening is about $7 (T_e/10^7$K$)^{1/2} \%$ per scatter.
If the abundance of Fe ionization stages with L-shell
vacancies is significant, resonance scattering can also occur.  In the
ionization stages with at least one L-shell electron, the K$\alpha$ photon
has a high probability of being lost as a result of resonance scattering
because the excited atom is more
likely to deexcite by Auger ionization than by reradiation.
The net result of all these processes is moderately strong limb-darkening, but
the degree of limb-darkening is a strong function of photon energy (Matt
et al. 1996).

   Thirdly, the relativistic effects must be evaluated (as in Matt et al.
1993 and Laor 1991).  There are several.
The photon energies and directions must be transformed from the rest frames of
the disk's fluid elements to the observer's frame.  In addition, if we
are considering regions not far outside the black hole's event horizon,
the photon trajectories suffer substantial bending by the gravitational field.
All these effects clearly depend strongly on the initial directions of the
photons, a fact which emphasizes the importance of a good calculation of
their fluid rest-frame angular distribution.

   Finally, each of these previous steps is implicitly a local calculation.
Each must be performed separately for different rings in the accretion disk,
and then added with weights appropriate to the run of emissivity with radius.
That distribution is, obviously, almost totally unconstrained {\it a priori},
so our procedure will usually be to guess some simple form, compute its
predictions, and compare to the data.

   To date, there is no model calculation in which all these elements
have been combined.  Perhaps the single most complete is that of Matt
et al. (1996), in which the ionization balance as a function of depth,
local angular distribution, and relativistic effects are all computed for two sample emissivity laws.  However, even in such a complex model there
are still unresolved issues that can be expected to change the
predicted profiles at the order unity level.  First, of course, is the
underlying disk model---they chose the standard Shakura-Sunyaev solution,
despite its instability.  Second, the ionizing radiation transfer is
treated in the diffusion approximation, despite the fact that few ionizing
photons penetrate to Thomson depths of more than $\sim 1$.  Finally,
their relativistic calculation is limited to the Schwarzschild
geometry, whereas realistic black holes are likely to be much better
described by the Kerr metric.  The line profiles and
angular distributions predicted by the two metrics are substantially different
when there is much line emission from within a few gravitational radii
(Laor 1991).  Thus, it is still a bit premature to expect quantitative
conclusions from fits of measured line profiles to disk emission models.

   Another line of approach may offer additional insight: variability analysis
in the spirit of reverberation mapping.  If the line emission
comes from a region comparable in size to the continuum source (as would
be predicted by the disk models), one of the key assumptions of ordinary
reverberation mapping---that the line-continuum delay
depends only on the position of the line-emitting material---is no longer
valid.  The continuum source distribution is also important because the
light travel times from different parts of the continuum source to a
particular part of the line source are significantly different.
Nonetheless, it is still possible to work out relationships between
light travel times and source sizes that are analogous to those used
in conventional reverberation mapping.  With adequate data, constraints
may then be put on the source size and geometry.

    The difficulties in carrying out this monitoring program have to do with
the very short timescales that are relevant.  If the Fe K$\alpha$ photons
are emitted from within 10 gravitational radii of the black hole, the light
travel times across the system are only $\sim 10^3 (M/10^7 M_{\odot})$~s.
Because X-ray variability power spectra are generally rather ``red", the
amplitudes of variations on such short timescales are quite small.  Even
the very large effective area detectors on {\it XTE} may provide only
marginal signal/noise on even the brightest sources.  The much smaller
effective area of {\it AXAF} at 6~keV will severely handicap its use
for such studies.  Moreover, the low Earth orbit in which many X-ray telescopes
are placed in order to avoid radiation backgrounds poses another problem:
most sources are occulted by the Earth for $\simeq 3000$~s out of every
5400~s.  A window function of this sort makes observations of variability
on 1000~s timescales extremely difficult.  Only if a sufficiently bright
source can be found near enough the pole of the orbit to be continuously in
view does this sort of experiment become feasible.

\section{Scorecard}

   From this quick survey of the achievements of AGN emission line studies
we may extract a list of how the different sub-areas have propelled us
along the path toward a fundamental understanding of AGN.  Ironically,
the greatest qualitative advances have come from two areas in which
efforts to date have been comparatively small-scale: maser spot kinematics,
and Fe K$\alpha$ spectroscopy.   Study of the former has yielded by far
the cleanest measurement of a central mass concentration in a galactic
nucleus, as well as the tightest upper bound on its size; work with
the latter technique has produced the strongest evidence we have for
the existence of relativistically deep potentials in AGN.  We may hope
for more quantitative results from Fe K$\alpha$ spectroscopy in the future,
but they will require improvements in both data and models.  A variety of
other uses of emission lines---maps of the region radiating the narrow
emission lines, spectropolarimetry revealing reflected emission lines---have
been of great aid in the construction of unification models for otherwise
apparently dissimilar classes of AGN.  Now that we realize that AGN
can look very different depending on the observer's viewing angle, the
presence of broad emission lines can be used as a crude indicator of
our angle relative to the system axis.  However,
despite the enormous amount of work lavished on studies of broad emission
line flux ratios, profiles, and variability, and the
substantial advances in our knowledge of how and where these lines are produced
that have come from these studies,
we have not gained from them any significant information bearing on the
most important questions of AGN research.  It remains to be seen whether
further efforts can break through the restraints imposed by the complexity
of their emission, and reach our true goals of solving the problems posed in
\S 1 of this review.

\end{document}